\documentclass[twocolumn,resetfootnote]{aastex701}

\usepackage{amsmath}
\usepackage{booktabs}

\newcommand{\eri}{EriIV-9808}
\newcommand{\cen}{CenI-5136}

\shorttitle{Chemical Abundances in Eridanus~IV and Centaurus~I}
\shortauthors{Heiger et al.}

\begin{document}

\title{Not-so-heavy metal(s): \\ Chemical Abundances in the Ultra-faint Dwarf Galaxies Eridanus~IV and Centaurus~I}

\correspondingauthor{Mair{\'e}ad E. Heiger}

\author[0000-0002-2446-8332]{Mair\'ead~E.~Heiger}
\affiliation{David A. Dunlap Department of Astronomy \& Astrophysics, University of Toronto, 50 St George Street, Toronto ON M5S 3H4, Canada}
\affiliation{Dunlap Institute for Astronomy \& Astrophysics, University of Toronto, 50 St George Street, Toronto, ON M5S 3H4, CA}
\email[show]{mairead.heiger@mail.utoronto.ca}

\author[0000-0002-4863-8842]{Alexander~P.~Ji}
\affiliation{Department of Astronomy \& Astrophysics, University of Chicago, 5640 S Ellis Avenue, Chicago, IL 60637, USA}
\affiliation{Kavli Institute for Cosmological Physics, University of Chicago, Chicago, IL 60637, USA}
\email{alexji@uchicago.edu}

\author[0000-0002-9110-6163]{Ting~S.~Li}
\affiliation{David A. Dunlap Department of Astronomy \& Astrophysics, University of Toronto, 50 St George Street, Toronto ON M5S 3H4, Canada}
\affiliation{Dunlap Institute for Astronomy \& Astrophysics, University of Toronto, 50 St George Street, Toronto, ON M5S 3H4, CA}
\email{ting.li@astro.utoronto.ca}

\author[0000-0002-4733-4994]{Joshua~D.~Simon}
\affiliation{Observatories of the Carnegie Institution for Science, 813 Santa Barbara St., Pasadena, CA 91101, USA}
\email{jsimon@carnegiescience.edu}
\author[0000-0002-9269-8287]{Guilherme Limberg}
\affiliation{Department of Astronomy \& Astrophysics, University of Chicago, 5640 S Ellis Avenue, Chicago, IL 60637, USA}
\affiliation{Kavli Institute for Cosmological Physics, University of Chicago, Chicago, IL 60637, USA}
\email{limberg@uchicago.edu}
\author[0000-0002-3690-105X]{Julio~A.~Carballo-Bello}
\affiliation{Instituto de Alta Investigaci\'on, Universidad de Tarapac\'a, Casilla 7D, Arica, Chile}
\email{jcarballo@academicos.uta.cl}

\author[0000-0003-1697-7062]{William~Cerny}
\affiliation{Department of Astronomy, Yale University, New Haven, CT 06520, USA}
\email{william.cerny@yale.edu}

\author[0000-0001-5143-1255]{Astha~Chaturvedi}
\affiliation{Department of Mathematics and Physics, University of Surrey, GU2 7XH, UK}
\email{aa07223@surrey.ac.uk}

\author[0000-0002-7155-679X]{Anirudh~Chiti}
\affiliation{Department of Astronomy \& Astrophysics, University of Chicago, 5640 S Ellis Avenue, Chicago, IL 60637, USA}
\affiliation{Kavli Institute for Cosmological Physics, University of Chicago, Chicago, IL 60637, USA}
\email{achiti@uchicago.edu}

\author[0000-0003-1680-1884]{Yumi~Choi}
\affiliation{NSF National Optical-Infrared Astronomy Research Laboratory, 950 North Cherry Avenue, Tucson, AZ 85719, USA}
\email{yumi.choi@noirlab.edua}

\author[0000-0002-1763-4128]{Denija~Crnojevi\'c}
\affiliation{Department of Physics and Astronomy, University of Tampa, 401 West Kennedy Boulevard, Tampa, FL 33606, USA}
\email{dcrnojevic@ut.edu}

\author[0000-0002-9144-7726]{Clara~E.~Mart\'inez-V\'azquez}
\affiliation{Gemini Observatory, NSF's NOIRLab, 670 N. A'ohoku Place, Hilo, HI 96720, USA}
\email{clara.martinez@noirlab.edu}

\author[0000-0003-0105-9576]{Gustavo~E.~Medina}
\affiliation{David A. Dunlap Department of Astronomy \& Astrophysics, University of Toronto, 50 St George Street, Toronto ON M5S 3H4, Canada}
\affiliation{Dunlap Institute for Astronomy \& Astrophysics, University of Toronto, 50 St George Street, Toronto, ON M5S 3H4, CA}
\email{gustavo.medina@utoronto.ca}

\author[0000-0001-9649-4815]{Burçin~Mutlu-Pakdil}
\affiliation{Department of Physics and Astronomy, Dartmouth College, Hanover, NH 03755, USA}
\email{Burcin.Mutlu-Pakdil@dartmouth.edu}

\author[0000-0001-9438-5228]{Mahdieh~Navabi}
\affiliation{Department of Mathematics and Physics, University of Surrey, GU2 7XH, UK}
\email{m.navabi@surrey.ac.uk}

\author[0000-0002-8282-469X]{Noelia~E.~D. No\"el}
\affiliation{Department of Mathematics and Physics, University of Surrey, GU2 7XH, UK}
\email{n.noel@surrey.ac.uk}

\author[0000-0002-6021-8760]{Andrew~B.~Pace}
\affiliation{Department of Astronomy, University of Virginia, 530 McCormick Road, Charlottesville, VA 22904, US}
\email{apace@virginia.edu}

\author[0000-0003-4479-1265]{Vinicius M. Placco}
\affiliation{NSF NOIRLab, Tucson, AZ 85719, USA}
\email{vinicius.placco@noirlab.edu}

\author[0000-0001-5805-5766]{Alexander H. Riley}
\affiliation{Institute for Computational Cosmology, Department of Physics, Durham University, South Road, Durham DH1 3LE, UK}
\email{alexander.riley2@durham.ac.uk}

\author[0000-0002-1594-1466]{Joanna~D.~Sakowska}
\affiliation{Instituto de Astrof\'isica de Andaluc\'ia, CSIC, Glorieta de la Astronom\'ia, E-18080 Granada, Spain}
\email{jsakowska@iaa.es}

\author[0000-0003-1479-3059]{Guy~S.~Stringfellow}
\affiliation{University of Colorado Boulder, Boulder, CO 80309, USA}
\email{guy.Stringfellow@colorado.edu}

\collaboration{all}{(DELVE Collaboration)}

\begin{abstract}

We present detailed chemical abundances of the brightest star in each of the ultra-faint dwarf galaxies Eridanus~IV and Centaurus~I using high-resolution Magellan/MIKE spectroscopy.
The brightest star in Centaurus~I, \cen, is a very metal-poor star with metallicity [Fe/H] = $-2.52\pm0.17$ and chemical abundances typical of a star in an ultra-faint dwarf galaxy.
We confirm that the star in Eridanus~IV, \eri, is extremely metal-poor ([Fe/H] = $-3.25\pm0.19$) and find that it is carbon-enhanced, with [C/Fe] = $1.07\pm0.34$, as is common for many stars at this metallicity.
Both stars are also neutron-capture deficient, which is typical of stars in ultra-faint dwarf galaxies, but less common in other environments.
We consider possible enrichment scenarios for \eri\ and tentatively conclude that it is unlikely to be the descendant of a single Pop III progenitor, despite its carbon-enhancement and low metallicity.

\end{abstract}

\section{Introduction}\label{sec:intro}

Ultra-faint dwarf galaxies (UFDs) lie at the low-mass extreme of galaxy formation and evolution. 
Conventionally defined as galaxies with $L\leq10^5L_{\odot}$ ($M_{V} > -7.7$), UFDs are the oldest, most chemically primitive galaxies known. 
They are also among the most dark-matter dominated systems yet observed, with mass-to-light ratios typically in the hundreds or thousands \citep[see][for a review]{simon_faintest_2019}. 
In concert, these characteristics have made UFDs versatile laboratories for science cases ranging from the nature of dark matter to the astrophysical sites of nucleosynthesis \citep[e.g.,][]{ackermann_searching_2015,ji_r-process_2016,simon_timing_2023}.

UFDs are sometimes called ``fossils", because they are thought to have formed before reionization and have long been quiescent \citep{bovill_pre-reionization_2009}. 
They host universally ancient, metal-poor stellar populations, perhaps best evinced by their extremely low mean metallicities ($-3\lesssim \mathrm{[Fe/H]}\lesssim -2$). 
UFDs are thought to experience only one or a few total star formation events \citep{frebel_chemical_2012}; as a result, their chemical enrichment is severely limited. 
The abundances of their stars should therefore encode the signatures of the earliest enrichment events, uncomplicated by generations of core collapse supernovae (CCSNe) or Type Ia supernovae (SNeIa) \citep{frebel_near-field_2015}. 
It is also plausible that UFDs contain stars enriched only by the first stars (which produce characteristic nucleosynthetic signatures), or even the first stars themselves, if any have survived to the present day \citep{frebel_chemical_2012,ji_preserving_2015,magg_predicting_2018}.

Metal-poor stars in the halo and in larger so-called classical dwarf spheroidal galaxies (dSph) have long been useful tools for characterizing enrichment events and studying nucleosynthetic yields
\cite[e.g.,][]{hill_first_2002, sivarani_first_2004, li_estimating_2014, choplin_intermediate_2021}.
The stars in UFDs specifically are notable for several reasons: more can be ascertained about their environment compared to halo stars, such as the star formation history
\citep[e.g.,][]{waller_cosmic_2023,chiti_detailed_2023}; they are, on average, less enriched than halo and dSph stars (and therefore are more likely to preserve the signatures of rare events); and stars in UFDs have unique abundance patterns compared to metal-poor halo and dSph stars, most notably a deficiency in neutron-capture elements \citep{simon_high-resolution_2010, koch_neutron-capture_2013, ji_chemical_2019}. 
This last point suggests that, although dSphs seem to be a continuous population and the line demarcating UFDs is somewhat arbitrary, UFDs seem to experience distinct evolutionary histories to their more massive counterparts.
Stellar chemical abundances are sensitive probes of the underlying galaxy physics like stochastic enrichment and feedback that is thought to become increasingly important in the low mass regime but remains poorly constrained \citep{carigi_chemical_2008, brown_quenching_2014, munshi_dancing_2019, agertz_edge_2020, applebaum_stochastically_2020, brauer_aeos_2025}.

The stars in UFDs are therefore a unique entryway to both stellar and galactic archaeology, making their detailed abundances extremely powerful.
Unfortunately, the same properties that make UFDs scientifically interesting---ancient stellar populations, low stellar mass---are exactly those that hamper measurement of their stars' abundances. 
High-resolution spectroscopy is necessary to measure detailed chemical abundances, but most stars in UFDs are too faint to perform high-resolution spectroscopy. 
As a result, both the number of UFDs with any high-resolution spectroscopy and the number of stars per UFD with high-resolution spectroscopy are small.
A little less than half of the $\sim$40 confirmed UFDs have any high-resolution observations; most have high-resolution spectra of $\lesssim$5 stars (e.g., \citealt{koch_highly_2008,frebel_segue_2014, ji_r-process_2016, ji_chemical_2016, chiti_chemical_2018, webber_chemical_2023, hansen_chemical_2024, zaremba_ghost_2025}; see also \citealt{pace_local_2024}).
Small samples preclude robust interpretation of observed abundance patterns, particularly the significance of any diversity or rarity.
Detailed abundances of stars from a larger sample of UFDs are thus extremely valuable to understanding chemical enrichment in the early universe.

Here, we present the first abundance analysis of the UFDs Eridanus~IV (Eri~IV) and Centaurus~I (Cen~I) using high-resolution Magellan/MIKE spectroscopy of their brightest known member star (\eri\ and \cen). 
Eri~IV and Cen~I were initially identified by \citet{cerny_eridanus_2021} and \citet{mau_two_2020} respectively, using the DECam Local Volume Explorer survey (DELVE), a photometric survey designed to study dwarf galaxies in the local volume \citep{drlica-wagner_decam_2021, drlica-wagner_decam_2022}.
They were subsequently followed up with medium-resolution Magellan/IMACS spectroscopy \citep{heiger_reading_2024} and deep Magellan/Megacam photometry \citep{casey_deep_2025}.
Both systems were found to have kinematic and orbital properties largely consistent with other UFDs, although Eri~IV has some unusual morphological features that are not yet fully understood (see \citealt{casey_deep_2025} for discussion).
Also, its observed metallicity distribution function appears to be right-skewed (left-leaning), the opposite of what is expected.
The cause (and significance) of this skew is also not yet fully understood \citep{heiger_reading_2024}.

In Section \ref{sec:data}, we describe the observations and data used in this work. 
Section \ref{sec:methods} details the methodology for determining stellar parameters, abundances, and uncertainties. 
We discuss the results of this analysis in Section \ref{sec:discussion}, including a comparison to the population of dSph (\ref{sec:comparison}), analysis of possible progenitors (\ref{sec:progenitors}), and interpretation of observed carbon abundance of \eri\ (\ref{sec:cemp}), and comment on the metallicity distribution function of Eri~IV (\ref{sec:mdf}).
In Section \ref{sec:conclusion}, we summarize these findings.

\section{Observations and Data Reduction} \label{sec:data}

\begin{figure*}
\centering
\includegraphics[width=0.8\textwidth]{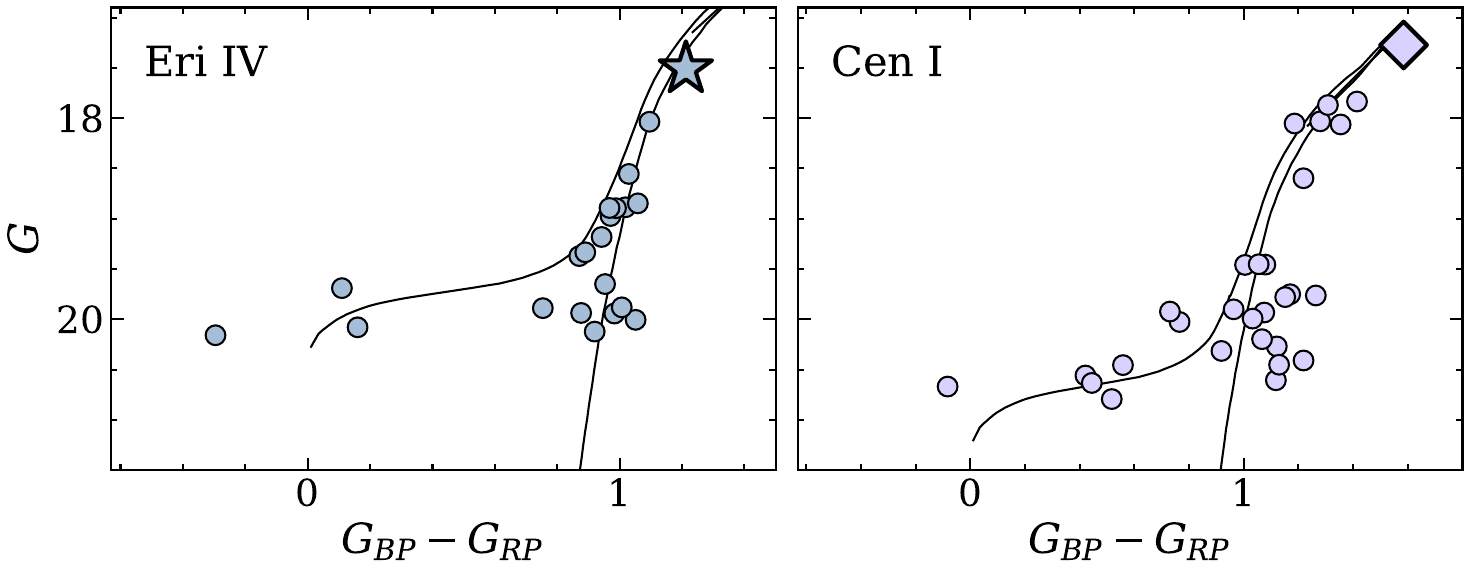}
\caption{Color-magnitude diagram in \textit{Gaia} $G$-band and $G_{BP} - G_{RP}$ of spectroscopically confirmed member stars in Eri~IV (left) and Cen~I (right) \citep{heiger_reading_2024}, with a 13 Gyr, [Fe/H]$=-2.2$ PARSEC isochrone \citep{bressan_parsec_2012,marigo_new_2017} overplotted. The targets of this follow-up, which are the brightest known member star in each galaxy, are marked by a star (\eri) and diamond (\cen).\label{fig:cmd}}
\end{figure*}

Using the high-probability member stars from \citet{cerny_eridanus_2021} and \citet{mau_two_2020}, we selected the brightest high-probability member star in each of Eri~IV (\eri) and in Cen~I (\cen) for high-resolution spectroscopic follow-up.
The color-magnitude diagram of the spectroscopically-confirmed member stars from \citet{heiger_reading_2024}, including \eri\ and \cen, are presented in Figure \ref{fig:cmd}, with our high-resolution targets highlighted.

The high-resolution spectra were collected using the Magellan Inamori Kyocera Echelle spectrograph (MIKE) on the 6.5m Magellan Clay telescope \citep{bernstein_mike_2003,bernstein_data_2015}. We used 2x2 binning and a 0.7\arcsec slit, which results in a resolution $R\sim35,000$ in the blue arm and $R\sim28,000$ in the red arm.

\eri\ was observed over two nights (2021/12/02-2021/12/03) with decent conditions (seeing $<$1\arcsec), for a total exposure time of 280min. \cen\ was observed on 2022/03/07 with decent conditions (seeing $<$1\arcsec); the exposure time was 160min. 
Individual exposures were 1800s-2400s, with ThAr lamp calibration frames taken every 1-2 exposures.
The data were reduced using \texttt{CarPy} \citep{kelson_evolution_2000, kelson_optimal_2003}. Details on the observations are presented in Table \ref{tab:observations}.

\begin{deluxetable*}{llrrccccccr}
\tablecaption{Summary of Observations\label{tab:observations}}
\tablehead{
\colhead{Star} & \colhead{\textit{Gaia} ID} & \colhead{R.A.} & \colhead{Decl.} & \colhead{$G$} & \colhead{$BP-RP$} &  \colhead{$t_{exposure}$} & \colhead{S/N} & \colhead{$v_{r}$}\\
 & & \colhead{(deg)} & \colhead{(deg)} & \colhead{(mag)} & \colhead{(mag)} & \colhead{(min)} & \colhead{4500/6500\:\AA} & \colhead{$\mathrm{\:km\:s^{-1}}$}}
\startdata
\eri & 3182722194000359808 & 76.418325 & $-$9.562303 & 17.51 & 1.21 & 280 & 18/42 & $-31.8\pm1.3$\\
\cen & 6146232547153275136 & 189.564404 & $-$40.934736 & 17.21 & 1.58 & 160 & 13/45& $~~50.0\pm1.4$ \\
\enddata
\end{deluxetable*}

\begin{figure*}
\centering
\includegraphics[width=\textwidth]{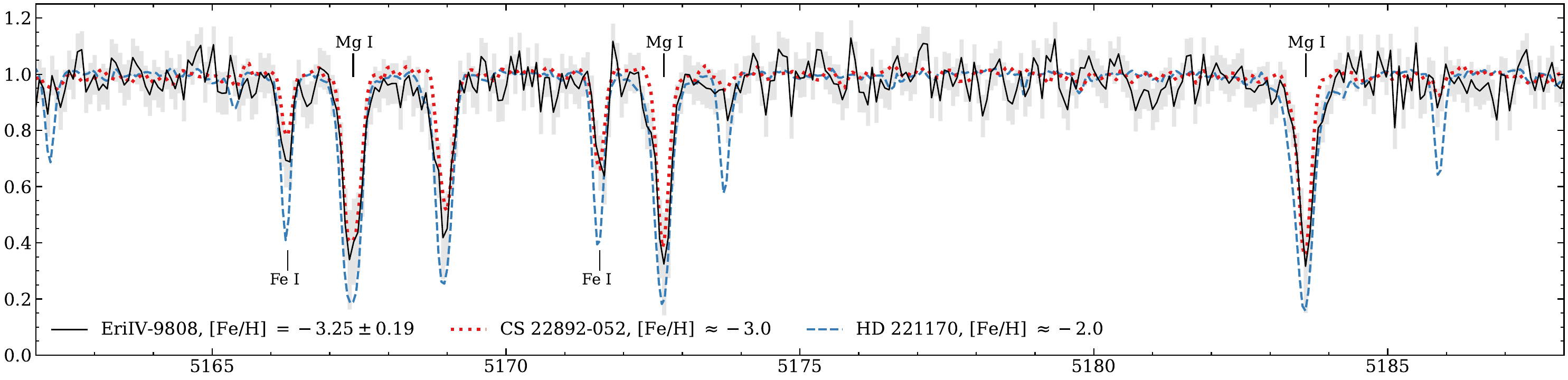}
\includegraphics[width=\textwidth]{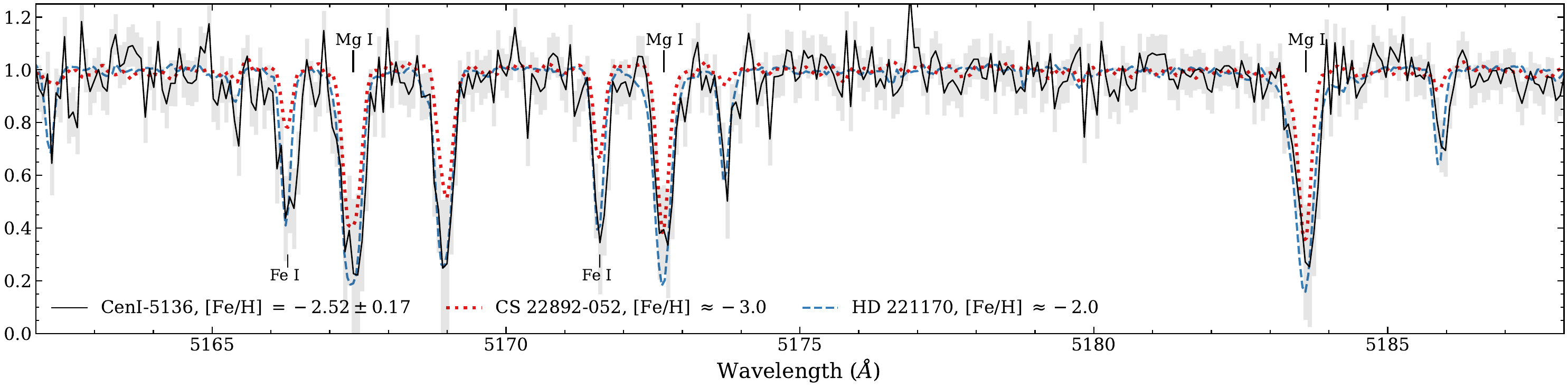}
\caption{Observed spectra (black lines and gray shaded region) around the Mg b triplet, a useful metallicity indicator. 
Spectra of the well-studied metal-poor stars HD 221170 (blue dashed line) \citep{burris_neutron-capture_2000,prugniel_database_2001} and CS22892-052 (red dotted line) \citep{sneden_extremely_2003} are plotted for comparison. 
(Top) This qualitative comparison indicates that \eri\ should be extremely metal-poor, which aligns well with the measured metallicity, [Fe/H] $=-3.25 \pm 0.19$.
(Bottom) Qualitative comparison of \cen\ suggests that its metallicity lies between these reference stars, especially when considering that \cen\ is cooler than both reference stars.
\label{fig:mgb}}
\end{figure*}

\section{Abundance Analysis} \label{sec:methods}

The spectral orders were Doppler-corrected, continuum normalized, and stitched using LESSPayne\footnote{\url{https://github.com/alexji/LESSPayne}} (Labeling Echelle Spectra with \texttt{SMHR} and \textit{Payne}) \citep{ji_lesspayne_2025}, which has been previously used to analyze MIKE spectra \citep[e.g.,][]{limberg_extending_2023, atzberger_chemical_2024}. 
\textit{The Payne} is a generative machine learning method for determining stellar parameters and abundances based on fitting synthetic models \citep{ting_payne_2019}. 
Here, \textit{The Payne} is used to automate the continuum normalization and to provide preliminary stellar parameter estimates via a full-spectrum fit. 
For the continuum normalization, prominent absorption features are masked, and then each order is fit with a cubic spline.
We show a portion of the normalized spectra in Figure \ref{fig:mgb}, as well as a comparison with other known metal-poor stars.

To measure the radial velocity, we first measure the radial velocity of each order with central wavelengths between 4000\AA\ and 6800\AA\ by cross-correlating the orders with the spectrum of HD122563. 
We iteratively sigma-clip orders with measured velocities more than 5$\sigma$ from the median velocity and apply a heliocentric velocity correction.
The resulting radial velocity is measured from 29 orders for \eri\ and 34 for \cen.
We report the inverse variance weighted average of the radial velocity measured from these orders in Table \ref{tab:observations}.
The weighted standard deviation of the velocities is added in quadrature with a 1$\mathrm{\:km\:s^{-1}}$ systematic error \citep{ji_southern_2020}.
The velocities are consistent within 1$\sigma$ with previous IMACS measurements \citep{heiger_reading_2024}, so there is no evidence of binarity for either star.

After normalization, we performed a standard 1D local thermodynamic equilibrium (LTE) analysis to determine stellar parameters and chemical abundances using the code \texttt{SMHR}\footnote{\url{https://github.com/andycasey/smhr}} \citep{casey_tale_2014}.
\texttt{SMHR} facilitates measurement of equivalent widths, determination of stellar parameters, and spectral synthesis.
For this analysis, we used $\alpha$-enhanced \texttt{ATLAS99} model atmospheres \citep{castelli_new_2003} and the 1D LTE radiative transfer code \texttt{MOOG} \citep{sneden_nitrogen_1973} including scattering \citep{sobeck_abundances_2011, sobeck_moog_scat_2023}.

\subsection{Determining stellar parameters} \label{sec:sp_meas}
\begin{deluxetable*}{llCCCC}
\tablecolumns{5} 
\tablecaption{Stellar parameters and uncertainties.\label{tab:sp}}
\tablehead{
\colhead{Star} & \colhead{Method} & \colhead{$T_{\mathrm{eff}}$} & \colhead{$\log{g}$} & \colhead{$\nu_{t}$} & \colhead{[M/H]} \\
 & & \colhead{(K)} & \colhead{(dex)} & \colhead{($\mathrm{km\:s}^{-1}$)} & \colhead{(dex)}
} 
\startdata
\eri & Photometric (fiducial) & 4672\pm90 & 1.32\pm0.22 & 2.47\pm0.24 & -3.25\pm0.19\\
\nodata & Spectroscopic & 4300\pm120 & 0.30\pm0.22 & 2.50\pm0.23 & -3.60\pm0.17 \\
\midrule
\cen & Photometric (fiducial) & 4127\pm87 & 0.44\pm0.21 & 2.89\pm0.24 & -2.52\pm0.17 \\
\nodata & Spectroscopic & 4168\pm119 & 0.10\pm0.21 & 2.96\pm0.26 & -2.65\pm0.14
\enddata
\end{deluxetable*}

In brief, we derive stellar parameters both photometrically and spectroscopically, then adopt the photometric parameters for the subsequent analysis.

The spectroscopic stellar parameters were determined using the method detailed in \cite{frebel_deriving_2013}.
After fitting Gaussian line profiles and measuring the equivalent widths of absorption features using \texttt{SMHR}, we used the Fe I and Fe II abundances as a function of excitation potential and reduced equivalent width to derive spectroscopic stellar parameters. 
Fe lines with reduced equivalent widths $>-4.5$ were excluded, as they are likely saturated.
Spectroscopic effective temperature ($T_{\mathrm{eff}}$) was determined by minimizing the slope of Fe I abundance as a function of excitation potential. 
Surface gravity ($\log{g}$) was determined by equilibrating Fe I and Fe II abundances. 
Microturbulence ($\nu_t$) was determined by minimizing the slope of Fe I abundance as a function of reduced equivalent width.
Fe II lines are usually preferred here, because microturbulence estimates are less affected by non-LTE effects when using Fe II compared to Fe I.
However, due to low metallicity and/or low signal-to-noise (S/N), the number of viable Fe II lines was too small for a reliable determination for these stars.
As such, we used Fe I lines instead.
Previous analysis comparing $\nu_t$ from Fe I versus Fe II suggests the bias this introduces is on the order of $+0.3 \mathrm{\:km\:s^{-1}}$ \citep{ji_southern_2020}.

To calculate spectroscopic stellar parameter uncertainties, we sum the random errors and systematic uncertainties in quadrature.
The random error of the effective temperature and microturbulence is estimated by varying the parameter to encompass the 1$\sigma$ error on the slope of Fe I or Fe II abundance as a function of excitation potential and reduced equivalent width respectively. 
The random error of $\log{g}$ is estimated by varying $\log{g}$ until the difference between Fe I and Fe II abundances matches the 1$\sigma$ standard error. 
The standard deviation of Fe I and Fe II abundances combined is used as the random error of the metallicity.
These random errors are added in quadrature with systematic uncertainties of $100$ K, $0.2$ dex, and $0.2~\mathrm{km\:s}^{-1}$ for $T_{\mathrm{eff}}$, $\log{g}$, and $\nu_{t}$, based on the estimates in \cite{usman_multiple_2024}. 
The systematic uncertainties help account for departures from 1D LTE.

We also determined stellar parameters from photometry. 
Temperature is derived using Equation 1 from \cite{mucciarelli_exploiting_2021}, using the coefficients for RGB stars and $BP-RP$ color. 
A reddening correction is applied to the \textit{Gaia} EDR3 photometry using $E(B-V)$ from \cite{schlegel_maps_1998} (as implemented in the \texttt{dustmap} module by \citealt{green_dustmaps_2018}) and extinction coefficients from \citet{gaia_collaboration_gaia_2018}. 
$\log{g}$ is determined using the following equation from \cite{venn_geminigraces_2017}:

\begin{multline}
    \log{g} = 4.44 + \log{M_{*}} + 4\log{\frac{T_{\mathrm{eff}}}{5780}} \\ + 0.4(G_{0} - \mu + BC(G) -  4.75)
\end{multline}

We assumed a stellar mass $M_{*} = 0.8\pm0.1~M_{\odot}$, typical for an old RGB star, and determined the bolometric correction $\mathrm{BC}(G)$ using the MIST (MESA Isochrones \& Stellar Tracks) bolometric correction tables \citep{dotter_mesa_2016,choi_mesa_2016}, wrapped by the \texttt{isochrones} package \citep{morton_isochrones_2015}. 
With these photometric parameters, we re-derived $\nu_{t}$ as above. 
We do this procedure first using the spectroscopic metallicity to calculate the effective temperature, and then repeat it using the resulting photometric metallicity from this first iteration to get our final photometric stellar parameters.

To calculate the uncertainty on the photometric effective temperature, we propagate the metallicity uncertainty and a magnitude uncertainty of 0.02 \citep[e.g.,][]{ji_southern_2020,usman_multiple_2024}.
This results in uncertainty in the effective temperature of $\pm35$K and $\pm25$K for \eri\ and \cen, respectively.
The systematic uncertainty for the photometric calibration reported by \citet{mucciarelli_exploiting_2021} is $\pm83$K.
We add these errors in quadrature to obtain the reported uncertainty on $T_{\textrm{eff}}$.
We propagate uncertainties in $T_{\textrm{eff}}$, $M_{*}$, $G_0$, and $\mu$ to determine the statistical uncertainties on $\log{g}$, which are $\pm0.08$ (\eri) and $\pm0.07$ (\cen) dex.
We add these in quadrature with a systematic error of 0.2 dex.
The uncertainty on the microturbulence and the metallicity are calculated the same way as for the spectroscopic stellar parameters.

The stellar parameters are tabulated in Table \ref{tab:sp}.
When comparing the values for \eri, we find the expected relation, which is that the spectroscopically-derived effective temperature, surface gravity, and metallicity are lower than the photometrically-derived values \citep{ezzeddine_r-process_2020}.
For \cen, we find that the spectroscopically-derived and photometrically-derived temperatures are actually very similar, and the spectroscopically-derived surface gravity and metallicity are lower as expected.
The profile of the Balmer lines is consistent with the low effective temperature derived with both methods. 
Compared to the previous calcium triplet metallicity measurements, the high-resolution metallicity measurements are lower \citep{heiger_reading_2024}.
For \eri, the photometric (spectroscopic) metallicity is consistent within 1 (2) $\sigma$ of the calcium triplet metallicity measurement.
For \cen, the photometric metallicity of \cen\ is consistent within 1.5 $\sigma$, but the spectroscopic metallicity is notably lower (consistent within 2.5 $\sigma$).

The discrepancy between spectroscopically and photometrically determined stellar parameters reflects non-LTE effects, as well as limitations of the models and uncertainties in the atomic data \citep{cayrel_first_2004,frebel_deriving_2013,ezzeddine_r-process_2020}. 
Spectroscopic stellar parameters are determined more self-consistently with chemical abundances, especially abundances measured by spectral synthesis, but they are determined with an LTE analysis of lines with known strong non-LTE effects (namely Fe I lines). 
Photometric stellar parameters are largely independent of such assumptions, but require empirical calibrations and reliable reddening.
Given the lack of availability of Fe II lines for both \eri\ and \cen, we consider the photometrically-derived stellar parameters to be more reliable and use them for the subsequent analysis.
The difference in [X/Fe] abundance when using photometric versus spectroscopic stellar parameters is typically $\sim0.2$ dex.
While the stellar parameters derived using these different methods are included for completeness, the choice of stellar parameters does not qualitatively affect the findings in this work.

\subsection{Determining chemical abundances} \label{sec:abun_meas}

To determine chemical abundances, we used \texttt{SMHR} to do a combination of curve-of-growth analyses and spectral syntheses. 
The curve of growth describes how the equivalent width of an absorption line changes with atomic abundance in a given stellar atmosphere; the abundance necessary to reproduce an observed equivalent width can therefore be calculated using a radiative transfer code and a stellar atmosphere model. 
Observed equivalent widths were measured by fitting Gaussian line profile to individual absorption features, with the exception of the Mg b triplet, for which Voigt profiles were used. 
The linelist is the same as in \cite{ji_southern_2020}.

\begin{figure*}
\centering
\includegraphics[width=\textwidth]{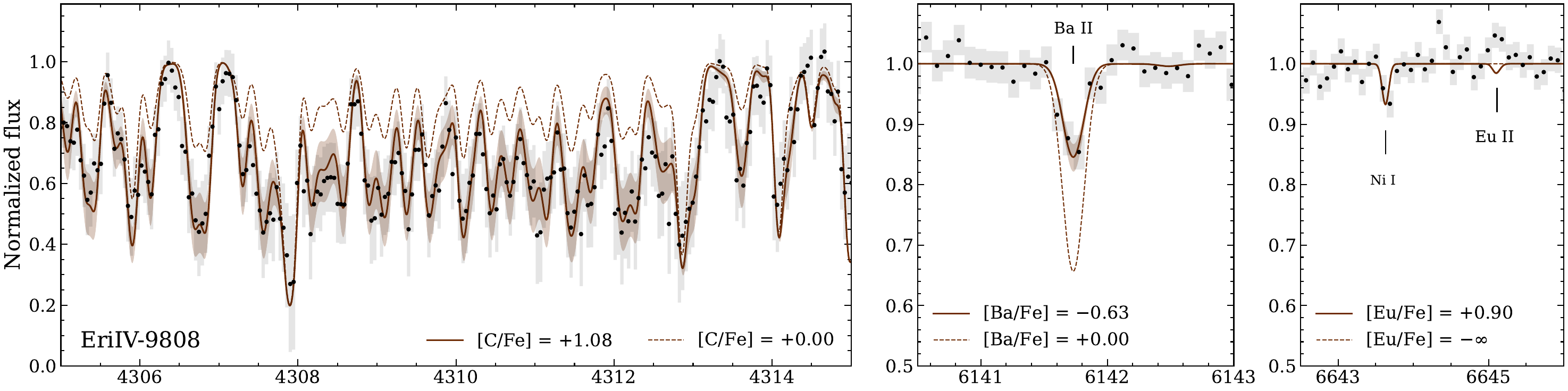}
\includegraphics[width=\textwidth]{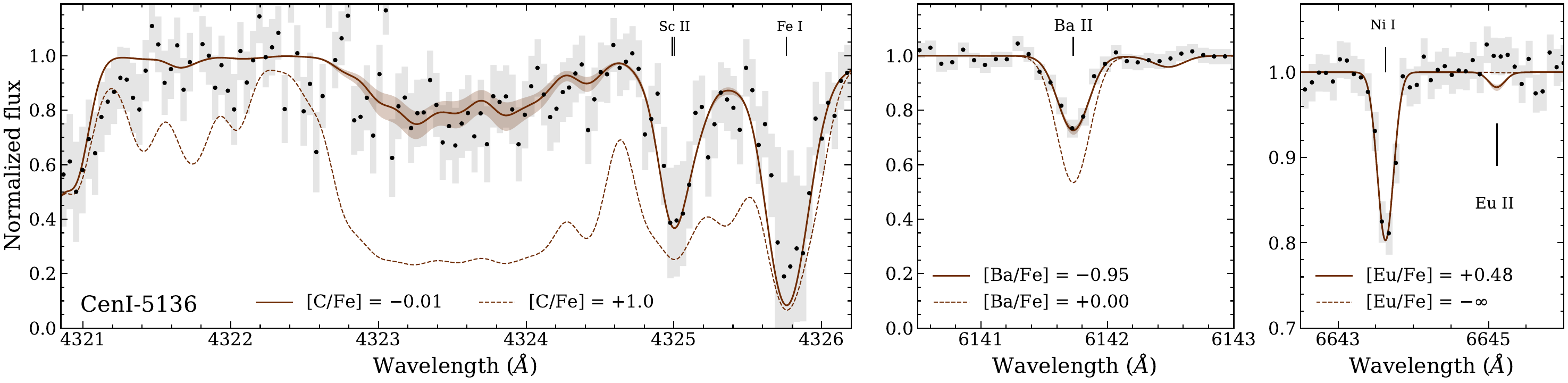}
\caption{Example regions of the observed spectra (black dots and shaded gray region) of \eri\ (top) and \cen\ (bottom). The best-fit spectral synthesis and 1$\sigma$ uncertainties are overplotted (solid brown line and shaded brown region).
For the Eu II region (right), the brown line is the 5$\sigma$ upper limit.
The dashed brown lines show what the spectra would look like with enhancement or deficiency of the highlighted abundances.
The selected CH absorption regions highlight that \eri\ shows strong enhancement, while \cen\ is clearly carbon-normal. 
The Ba and Eu absorption features demonstrate that both \eri\ and \cen\ are deficient in neutron-capture elements.
\label{fig:syn_spec}}
\end{figure*}

Molecular features, heavily blended lines, and lines subject to hyperfine splitting are poorly modeled with this method. 
We instead used spectral synthesis to measure abundances from such features (including C-H, C-N, Al I, Sc II, Mn I, Ni I, and Ba II), and to measure 5$\sigma$ upper limits for V I and Eu II. 
The measured abundance of each line for both stars is tabulated in Table \ref{tab:lines}; the average abundances are reported in Tables \ref{tab:e4_abs} for \eri\ and \ref{tab:c1_abs} for \cen.
We show several key regions of the spectra and the best-fit spectral synthesis in Figure \ref{fig:syn_spec}. 
Different regions are shown for each star, due to metallicity and S/N constraints that affect the quality of the fits.

Here we note the major features used to derive each abundance and comment on specific measurements:

\textit{Light elements (C, N, O):}\label{sec:carbon}
C is measured with spectral synthesis of two molecular C-H features for \eri\ and one for \cen\ (due to low S/N). 
We applied a correction to the C abundance for evolutionary stage as per \cite{placco_carbon-enhanced_2014}\footnote{\url{https://vplacco.pythonanywhere.com/}}, which is $+0.64$ dex for \eri\ and $+0.76$ dex for \cen\ (assuming photometrically derived stellar parameters).
The [C/H] and [C/Fe] values reported in Tables \ref{tab:e4_abs} and \ref{tab:c1_abs} are corrected; $\log{\epsilon}$ is not. 
The individual line measurements in Table \ref{tab:lines} are also not corrected.
Measurement of O from the forbidden line at 6300 \AA\ was attempted but unsuccessful.
Measurements of N from the CN band at $\sim$3865-3890 \AA\ were also attempted, but low S/N in the blue prevented reliable measurements for both \eri\ (S/N$\sim$5) and \cen\ (S/N$\sim$3). 

\textit{$\alpha$-elements (Mg, Si, Ca, and Ti):}\label{sec:alpha}
The $\alpha$-elements are measured from a number of features across a wide wavelength range for both stars; which features are usable varies due to the different S/N and metallicity of \eri\ and \cen.

\textit{Odd-Z elements (Na, Al, K, and Sc):}\label{sec:odd_z}
Na is derived from two features at 5890 \AA\ and 5896 \AA.
Al is measured only for \eri, from the feature at 3962 \AA\ using spectral synthesis.
In \cen, the S/N is too low to reliably use this line (S/N $\sim 4$).
We also note that although we report a measurement for \eri, it should be regarded with caution.
This Al feature is blended and subject to substantial non-LTE effects; it also lies in the blue, where the S/N is lowest.

K is measured from resonant lines at 7665 \AA\ and 7699 \AA.
Fortunately, these features are not blended with telluric absorption for either star.
A handful of Sc features are measured in both stars with spectral synthesis due to blending and hyperfine structure. 
The bluest lines are not measured for \cen\, again due to low S/N.

We do not apply any non-LTE corrections, but depending on the stellar parameters, Na, Al, and K lines can have meaningful departures from LTE.
For Al, the correction is expected to be positive and large, on the order of several tenths of a dex up to about +1 dex \citep[e.g.,][]{mashonkina_formation_2017, nordlander_non-lte_2017}.
The corrections for Na and K are typically less significant and negative for stars like \eri\ or \cen\ \citep{andrievsky_non-lte_2010,mashonkina_1d_2023}.

\textit{Iron-peak elements (V, Cr, Mn, Fe, Ni, and Zn):}\label{sec:iron_peak}
Fe I and Fe II abundances are measured from many lines across a wide wavelength range for both stars.
In \cen, we measure V and Zn from one feature each (at 4384 \AA\ and 4811 \AA\ respectively). 
The V feature is subject to heavy blending, so we use spectral synthesis.
For \eri, we only obtain an upper limit for V and make no measurement of Zn, as it is substantially more metal-poor.

Cr and Ni are measured from eight and six features respectively in \cen. 
In \eri, we measure Cr at only 5206 \AA\ and Ni at 5477 \AA\, again due to its low metallicity.
An upper limit on Mn is measured using features at 4762 and 4766 (\cen\ only) \AA.

\textit{Neutron-capture elements (Sr, Ba, Eu):}\label{sec:ncap}
Sr is measured from the lines at 4077 \AA\ (\eri\ only) and 4215 \AA.
Ba is measured with spectral synthesis from 3-5 lines.
Measurement of Eu is attempted, but we obtain only upper limits using spectral synthesis of the feature at 4205 \AA.

We calculate uncertainties for each line as in \cite{atzberger_chemical_2024}.
Briefly, the uncertainty on the abundance measured from a particular line includes a statistical uncertainty from the equivalent width fit, uncertainty due to stellar parameter uncertainties (the dominant source of uncertainty), and an additional systematic error of 0.1 dex. 
This 0.1 dex systematic floor helps account for uncertainties in, e.g., the model atmospheres or atomic data. 
The uncertainty associated with a stellar parameter, e.g., $e_{T_{\mathrm{eff}}}$, is the difference in abundance measured from that line at the 1$\sigma$ value of the parameter. 
Each of these sources of uncertainty are assumed to be independent, so the total uncertainty on each line measurement is therefore:

\begin{equation}
\begin{split}
    \sigma_{i}^{2} = e_{T_{\mathrm{eff},i}}^2 + e_{\log{g},i}^2 + e_{\mathrm{[M/H]},i}^2 + e_{{\nu_{t}},i}^2+\sigma_{stat}^2 + \sigma_{sys}^{2}
\end{split}
\end{equation}

\begin{deluxetable*}{lccccccccccc}
\tabletypesize{\footnotesize}
\tablecaption{Line measurements \label{tab:lines}}
\tablehead{
\colhead{Star} & 
\colhead{Wavelength} & \colhead{Element} & \colhead{Species} & \colhead{$\log{(\epsilon)}$} & \colhead{$\sigma_{i}$} & \colhead{$\sigma_{stat}$} & \colhead{$e_{T_{\mathrm{eff}}}$} & \colhead{$e_{\log{g}}$} & \colhead{$e_{\nu_{t}}$} & \colhead{$e_{\mathrm{[M/H]}}$} & \colhead{$e_{sys}$}} 
\startdata
EriIV\_9808 & 3905.52 & Si I & 14.0 & 4.38 & 0.35 & 0.29 & 0.10 & -0.07 & -0.12 & -0.04 & 0.10 \\
EriIV\_9808 & 3961.52 & Al I & 13.0 & 1.78 & 0.25 & 0.2 & 0.11 & -0.02 & -0.05 & -0.02 & 0.10 \\
EriIV\_9808 & 4001.66 & Fe I & 26.0 & 4.44 & 0.23 & 0.16 & 0.12 & -0.00 & -0.01 & 0.00 & 0.10 \\
\nodata & \nodata & \nodata & \nodata & \nodata & \nodata & \nodata & \nodata & \nodata & \nodata & \nodata \\
\enddata
\tablecomments{This table is available in its entirety in machine-readable format.}
\end{deluxetable*}

The results for individual line measurements are tabulated in Table \ref{tab:lines}. 
Using these measurements, we compute a weighted average of each abundance as in \cite{atzberger_chemical_2024}; these abundances are reported in Tables \ref{tab:e4_abs} and \ref{tab:c1_abs}. 
The solar scaled abundances are computed using the solar photospheric abundances from \citet{asplund_chemical_2009}.

\section{Discussion}\label{sec:discussion}

\subsection{Comparison to other UFDs}\label{sec:comparison}
\begin{deluxetable}{cccccccccc}
\tabletypesize{\footnotesize}
\tablecaption{Abundance measurements for \eri\label{tab:e4_abs}}
\tablehead{
\colhead{Element} & \colhead{Species} & \colhead{N} & \colhead{$\log{(\epsilon)}$} & \colhead{[X/H]} & \colhead{[X/Fe]} & \colhead{$\sigma_{\textrm{[X/H]}}$} & \colhead{$\sigma_{\textrm{[X/Fe]}}$} \\
& & & \colhead{(dex)} & \colhead{(dex)} & \colhead{(dex)} & \colhead{(dex)} & \colhead{(dex)}}
\startdata
\midrule
C-H & 106.0 & 2 & 5.69 & -2.18\tiny{\tablenotemark{a}} & 1.07\tiny{\tablenotemark{a}} & 0.33 & 0.34 \\
Na I & 11.0 & 2 & 3.44 & -2.8 & 0.45 & 0.24 & 0.29 \\
Mg I & 12.0 & 5 & 5.14 & -2.46 & 0.79 & 0.14 & 0.25 \\
Al I & 13.0 & 1 & 1.78 & -4.67 & -1.42 & 0.28 & 0.34 \\
Si I & 14.0 & 2 & 4.71 & -2.8 & 0.45 & 0.27 & 0.33 \\
K I & 19.0 & 2 & 2.39 & -2.64 & 0.61 & 0.23 & 0.31 \\
Ca I & 20.0 & 5 & 3.33 & -3.01 & 0.24 & 0.14 & 0.25 \\
Sc II & 21.1 & 4 & -0.16 & -3.31 & -0.06 & 0.22 & 0.32 \\
Ti I & 22.0 & 2 & 2.08 & -2.87 & 0.38 & 0.19 & 0.27 \\
Ti II & 22.1 & 9 & 2.07 & -2.88 & 0.37 & 0.16 & 0.28 \\
V I & 23.0 & 1 & $<$1.98 & $<$-1.95 & $<$1.3 & 0.22 & 0.28 \\
Cr I & 24.0 & 1 & 1.81 & -3.83 & -0.58 & 0.23 & 0.29 \\
Mn I & 25.0 & 1 & $<$2.62 & $<$-2.81 & $<$0.44 & 0.14 & 0.26 \\
Fe I & 26.0 & 60 & 4.25 & -3.25 & 0.0 & 0.25 & 0.31 \\
Fe II & 26.1 & 9 & 4.21 & -3.29 & -0.04 & 0.20 & 0.32 \\
Ni I & 28.0 & 1 & 3.06 & -3.16 & 0.09 & 0.22 & 0.29 \\
Sr II & 38.1 & 2 & -1.51 & -4.38 & -1.13 & 0.25 & 0.32 \\
Ba II & 56.1 & 3 & -1.7 & -3.88 & -0.63 & 0.19 & 0.29 \\
Eu II & 63.1 & 1 & $<$-2.08 & $<$-2.6 & $<$0.65 & 0.16 & 0.32 \\
\bottomrule
\enddata
\tablenotetext{a}{Corrected for evolutionary stage per \citet{placco_carbon-enhanced_2014}}
\end{deluxetable}

\begin{deluxetable}{cccccccccc}
\tabletypesize{\footnotesize}
\tablecaption{Abundance measurements for \cen\label{tab:c1_abs}}
\tablehead{
\colhead{Element} & \colhead{Species} & \colhead{N} & \colhead{$\log{(\epsilon)}$} & \colhead{[X/H]} & \colhead{[X/Fe]} & \colhead{$\sigma_{\textrm{[X/H]}}$} & \colhead{$\sigma_{\textrm{[X/Fe]}}$} \\
& & & \colhead{(dex)} & \colhead{(dex)} & \colhead{(dex)} & \colhead{(dex)} & \colhead{(dex)}}
\startdata
\midrule
C-H & 106.0 & 1 & 4.88 & -2.78\tiny{\tablenotemark{a}} & -0.01\tiny{\tablenotemark{a}} & 0.22 & 0.28 \\
Na I & 11.0 & 2 & 3.25 & -2.99 & -0.22 & 0.32 & 0.27 \\
Mg I & 12.0 & 4 & 5.42 & -2.18 & 0.59 & 0.19 & 0.18 \\
Si I & 14.0 & 2 & 5.75\tiny{\tablenotemark{b}} & -1.76\tiny{\tablenotemark{b}} & 1.01\tiny{\tablenotemark{b}} & 0.22 & 0.27 \\
K I & 19.0 & 1 & 2.96 & -2.07 & 0.7 & 0.25 & 0.22 \\
Ca I & 20.0 & 12 & 3.88 & -2.46 & 0.31 & 0.20 & 0.20 \\
Sc II & 21.1 & 2 & 0.64 & -2.51 & 0.26 & 0.17 & 0.25 \\
Ti I & 22.0 & 8 & 2.04 & -2.91 & -0.14 & 0.21 & 0.16 \\
Ti II & 22.1 & 14 & 2.58 & -2.37 & 0.4 & 0.14 & 0.24 \\
V I & 23.0 & 1 & 0.63 & -3.3 & -0.53 & 0.37 & 0.34 \\
Cr I & 24.0 & 8 & 2.51 & -3.13 & -0.36 & 0.24 & 0.20 \\
Mn I & 25.0 & 2 & $<$2.3 & $<$-3.13 & $<$-0.36 & 0.15 & 0.19 \\
Fe I & 26.0 & 33 & 4.73 & -2.77 & 0.0 & 0.18 & 0.16 \\
Fe II & 26.1 & 5 & 5.0 & -2.5 & 0.27 & 0.23 & 0.32 \\
Ni I & 28.0 & 6 & 3.26 & -2.96 & -0.19 & 0.17 & 0.15 \\
Zn I & 30.0 & 1 & 1.91 & -2.65 & 0.12 & 0.18 & 0.28 \\
Sr II & 38.1 & 1 & -1.21 & -4.08 & -1.31 & 0.47 & 0.49 \\
Ba II & 56.1 & 5 & -1.54 & -3.72 & -0.95 & 0.16 & 0.24 \\
Eu II & 63.1 & 1 & $<$-1.77 & $<$-2.29 & $<$0.48 & 0.17 & 0.27 \\
\bottomrule
\enddata
\tablenotetext{a}{Corrected for evolutionary stage per \citet{placco_carbon-enhanced_2014}}
\tablenotetext{b}{Large uncertainties despite detection/measurement of the feature. Caution is advised when interpreting this abundance.}
\end{deluxetable}

\begin{figure*}[!p]
\centering
\includegraphics[width=\textwidth]{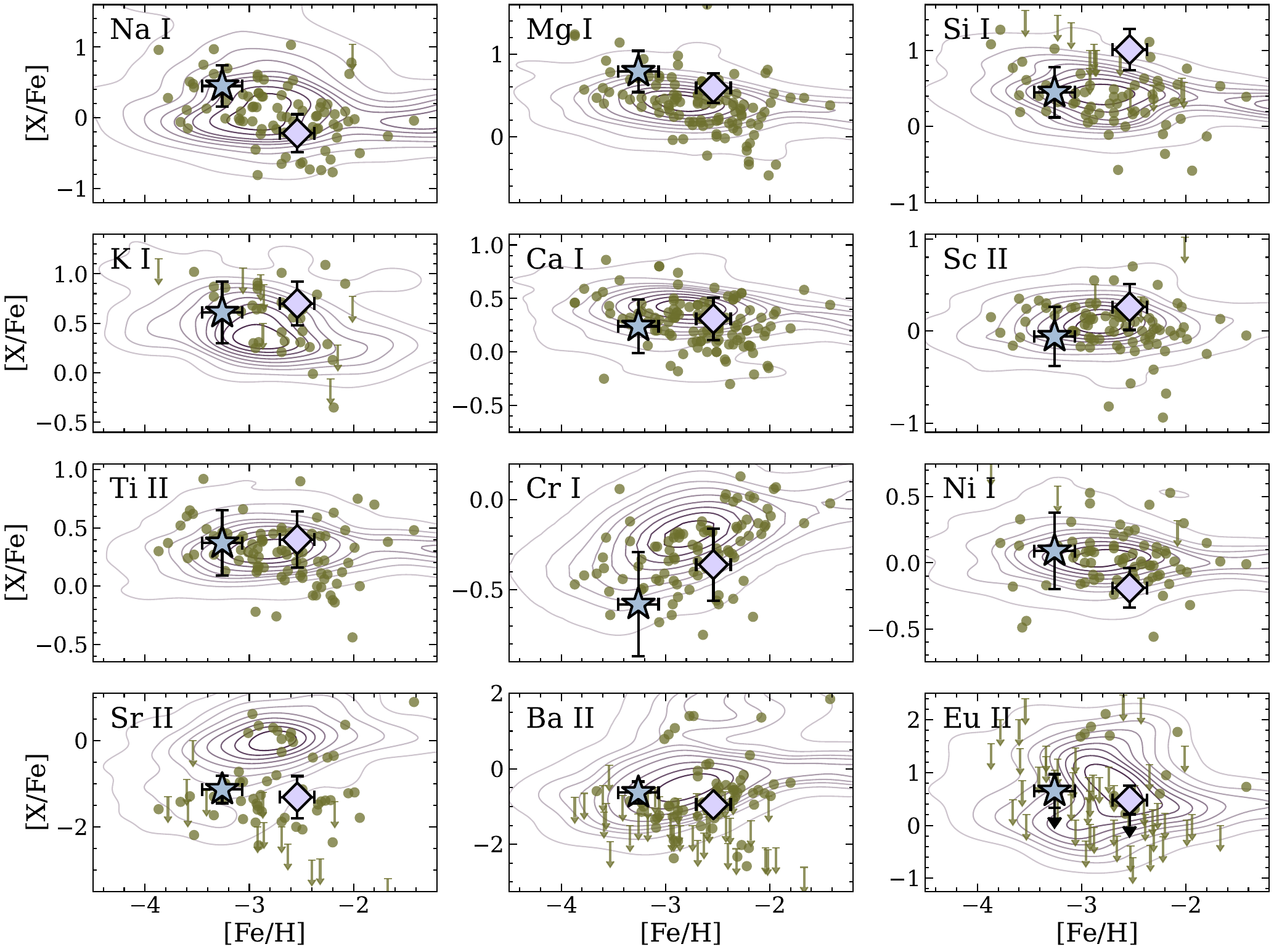}
\caption{
[X/Fe]-[Fe/H] for \eri\ (blue star) and \cen\ (purple diamond), plotted with a sample of stars in UFDs (green points) and the stellar halo (contour lines).
With a few exceptions, \eri\ and \cen\ follow halo and/or UFD trends for most chemical abundances. 
Data for the UFD sample are from: \cite{koch_highly_2008,feltzing_evidence_2009,frebel_high-resolution_2010,simon_high-resolution_2010,norris_chemical_2010,lai_feh_2011,gilmore_elemental_2013,koch_neutron-capture_2013,ishigaki_chemical_2014,frebel_segue_2014,ji_chemical_2016,ji_r-process_2016,roederer_diverse_2016,hansen_r-process_2017,kirby_triangulum_2017,nagasawa_chemical_2018,chiti_chemical_2018,spite_cemp-no_2018,marshall_chemical_2019,hansen_chemical_2020,chiti_magellanimacs_2022,waller_cosmic_2023}.
Data for the halo sample are from:
\cite{aoki_high-resolution_2013,cohen_normal_2013,norris_most_2013, roederer_search_2014, jacobson_high-resolution_2015,lombardo_chemical_2022,li_four-hundred_2022, nissen_abundances_2024}.
\label{fig:ab_grid}}
\end{figure*}

In Figure \ref{fig:ab_grid}, we plot a selection of the chemical abundances of \eri\ and \cen\ compared to those of stars in other UFDs and in the stellar halo.
Both stars generally follow the expected trends for stars in UFDs at a given metallicity for $\alpha$-elements (Mg, Si, Ca, Ti) and iron-peak elements (Ti, Cr, Ni). 
We note that the Si abundance in \cen\ is higher than expected for a star of this metallicity, but it is measured from only two lines in the blue, so we caution against over-interpreting this measurement.
The [$\alpha$/Fe] ratios suggests that both stars likely formed before dominant enrichment by Type Ia supernovae (which produce iron-peak elements and therefore tend to decrease the [$\alpha$/Fe] ratio) \citep{nomoto_nucleosynthesis_2013}. 
The odd-Z element abundances (produced in CCSNe) are also generally consistent with expected trends.

We also highlight that both \eri\ and \cen\ are neutron-capture deficient, with substantially sub-solar [Sr, Ba/Fe] ratios and only an upper limit on [Eu/Fe]. 
The center and right panels of Figure \ref{fig:syn_spec} highlight this neutron-capture deficiency and show what the spectra would look like if they were in fact enriched in Ba II and Eu II.
Low neutron-capture element abundance is characteristic of stars formed in UFDs, which tend to be neutron-capture deficient even compared to halo stars or stars in larger dSphs at comparable metallicities \citep{simon_high-resolution_2010,koch_neutron-capture_2013,ji_chemical_2019}.
This deficiency is most likely attributed to stochastic enrichment, wherein a poorly-sampled initial mass function results in a stellar population that experiences no or very few neutron-capture enrichment events \citep{ji_chemical_2019}. 
Some neutron-capture enrichment sources like asymptotic giant branch (AGB) stars (the primary producers of Sr and Ba) are also delayed.
In UFDs like Eri~IV and Cen~I, star formation is thought to be less extended, due to their vulnerability to supernova feedback and reionization \citep{brown_quenching_2014,weisz_star_2014, agertz_edge_2020}.
As such, ubiquitous neutron-capture deficiency could arise if star formation ceases before these delayed sources can contribute processed material to the ISM. 

\subsection{Progenitors}\label{sec:progenitors}

\begin{figure*}
\centering
\includegraphics[width=\textwidth]{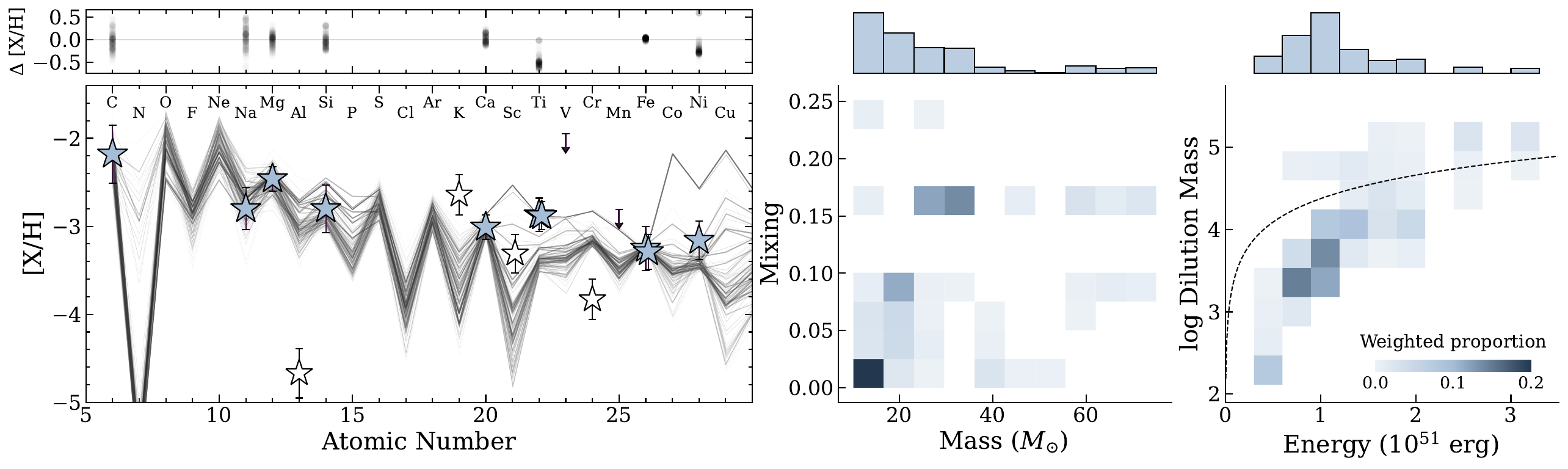}
\caption{(Left) Fits to the metal-free SNe models by \cite{heger_nucleosynthesis_2010} (transparent lines) for \eri, and the residuals. Abundances included in the fit are opaque blue stars; abundances excluded due to non-LTE effects or model uncertainties are white stars. Darker, thicker lines represent models with smaller $\chi^{2}$ values. Models within $5\sigma$ of the minimum $\chi^{2}$ value (which corresponds to $\chi^{2}_{\nu} \lesssim 15$; 148 models) are visualized here.
(Center) Weighted histogram of the mass and mixing parameters of the best-fitting models. 
(Right) Weighted histogram of the dilution mass and energy of the best-fitting models. The dashed black line shows approximately the mass swept up by a single SN remnant expanding into a low-metallicity interstellar medium \citep{ryan_extremely_1998}.
\label{fig:prog}}
\end{figure*}

With [Fe/H]$=-3.25\pm0.19$, \eri\ is sufficiently metal-poor to consider enrichment by only one supernova progenitor. 
To identify possible progenitors, we fit the theoretical 1D SNe yield models from \citet{heger_nucleosynthesis_2010} using the $\chi^2$ minimization routine implemented in \texttt{starfit}\footnote{\url{https://starfit.org}}.
The 16,800 models cover a range of masses ($9.6-100~M_{\odot}$), energies ($0.3-10\times10^{51}~\mathrm{erg}$), and mixing of the ejecta ($0.0-0.25$). 
We restrict the fit to elements lighter than Zn (Z $<$ 30) and exclude Al, K, Sc, and Cr from the fits due to the large and uncertain non-LTE effects and/or known model uncertainties \citep{heger_nucleosynthesis_2010}.
The fits are shown in Figure \ref{fig:prog}. 
Measured abundances included in the fit are opaque blue stars; abundances excluded are semi-transparent white stars.
The best-fit progenitor is a low mass ($\sim13.4~M_{\odot}$), normal energy ($\sim0.9\times10^{51}~\mathrm{ erg}$) supernova with minimal mixing.
The constraint is not strong, however, as none of the models is an especially good fit, with reduced $\chi^{2}\gtrsim5$.
A wide range of parameters is possible, as shown in the histograms in Figure \ref{fig:prog}, although generally a normal energy, lower mass progenitor ($M\lesssim20~M_{\odot}$) is preferred.

Ti and Ni in particular are systematically underestimated by the majority of the best-fit models (as demonstrated in the residual plot in Figure \ref{fig:prog}; see also \citealt{rossi_hidden_2024}).
The models that best reproduce the observed Ti abundance overestimate Ni by approximately $+0.5~\mathrm{dex}$.
This issue is reflected in the reduced $\chi^{2}$ values, which suggest that the models inadequately describe the data.
These systematic issues could reflect missing physics in the yield models, or simply that a single progenitor is not sufficient to reproduce the observed abundances.
Also, the majority of the best-fit models all predict extremely low N abundance.
This is a known artifact that arises when the measured N abundance is missing (or only an upper limit is available) that may bias the preferred progenitor mass to lower values \citep{placco_metal-poor_2015}.
If we consider only those models with [N/Fe] $\gtrsim -1.0$, for example, the best-fit progenitors are clustered at $20-30~M_{\odot}$ and $60-70~M_{\odot}$, rather than $<20~M_{\odot}$. 

We also note that the dilution mass for most of the best-fitting models is inconsistent with the energy of the SN.
The dilution mass, a free parameter, is the mass of hydrogen into which the ejecta are mixed to dilute them to the observed abundances.
The dashed line in Figure \ref{fig:prog} shows the mass expected to be swept up by a SN of a given energy in typical conditions; this line is essentially a minimum allowable dilution mass \citep{ryan_extremely_1998}. 
Further dilution (models above the line) can occur due to turbulent mixing \citep{ji_detailed_2020}, but less dilution is physically inconsistent with the energy of the SN.
Essentially, models with low dilution mass eject too little metal mass to reach good agreement with the observed abundances of \eri\ unless the ejecta are mixed with unphysically low masses of hydrogen.
This suggests that the gas out of which \eri\ formed may have been enriched by multiple SNe, rather than one.

We repeat this procedure with \cen, but predictably, none of the theoretical models is a good match. 
While there are metal-free SNe that can enrich to high metallicities, like pair instability supernovae or hypernovae, these have additional signatures not seen in the abundances of \cen. 
\cen\ is much more likely to be enriched by multiple progenitors than a single metal-free supernova.

\subsection{\eri\ is a carbon-enhanced metal-poor star}\label{sec:cemp}

\begin{figure*}
    \includegraphics[width=\textwidth]{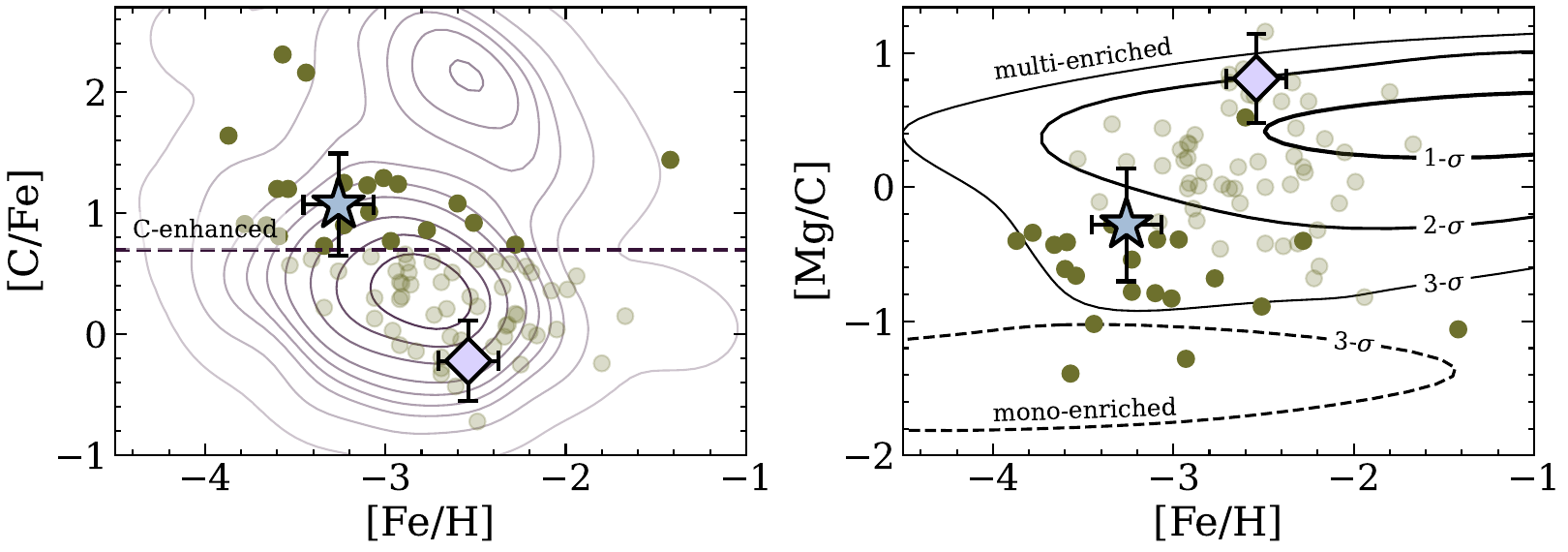}
    \caption{(Left) [C/Fe]-[Fe/H] for the same sample of stars in UFDs (green points) and the stellar halo (contour lines) as Figure \ref{fig:ab_grid}. For the literature sample, carbon-enhanced stars are opaque; carbon-normal stars are transparent. \eri\ is carbon-enhanced. \cen\ is carbon-normal. Note that an evolutionary correction to the C abundance has not been uniformly applied in the UFD and literature samples, so some of the stars considered carbon-normal may be carbon-enhanced, especially at lower metallicities.
    (Right) [Mg/C]-[Fe/H]. Also shown are the approximate contours denoting the distribution of multi- and mono-enriched stars in \citet{hartwig_descendants_2018}. \eri\ and many of the CEMP stars in UFDs fall in an ambiguous or perhaps transitional region in the tails of the multi-enriched distribution. This [Mg/C] region is suggested by \cite{rossi_understanding_2023} to reflect enrichment by Pop III SNe \label{fig:c_fe}}
\end{figure*}

With [C/Fe] = $1.07\pm0.34$, \eri\ is a carbon-enhanced metal-poor (CEMP) star. 
Conventionally defined as [C/Fe]$>0.7$ \citep{aoki_carbon-enhanced_2007}, carbon enhancement is common among metal-poor stars, especially at the metallicity of \eri, as Figure \ref{fig:c_fe} shows. 
Approximately 30-70\% of stars with [Fe/H] $\lesssim-3.0$ are carbon enhanced \citep{arentsen_inconsistency_2022}.
The designation of \eri\ as carbon-enhanced is not affected by the choice of stellar parameters.

Conventionally, \eri\ is considered a CEMP--no star, as in `no neutron-capture enrichment' (versus CEMP--$s/i$, which show slow or intermediate neutron-capture process enrichment) \citep{beers_discovery_2005,aoki_carbon-enhanced_2007, hampel_intermediate_2016}. 
CEMP--no stars are thought to be intrinsically enriched, in that their abundances reflect their natal environment.
Carbon enhancement in metal-poor stars can be due to pollution by an AGB companion, but such extrinsic enrichment also produces $s$-process enhancement atypical of CEMP stars in UFDs, including \eri\ \citep{masseron_holistic_2010}. 

Intrinsic carbon enhancement is interesting in metal-poor stars for its potential importance in gas cooling at extremely low metallicities and as a possible signature of enrichment by the first stars (Pop III stars) \citep{bromm_formation_2003, frebel_probing_2007,norris_most_2013, marassi_origin_2014, placco_observational_2016,chiaki_classification_2017}.
\cite{ji_preserving_2015} found that carbon enhancement by first-generation stars is quickly erased by only a few normal second-generation CCSNe, suggesting that if we assume the carbon enhancement of CEMP--no stars is indeed due to Pop III enrichment, they can be considered `true' second stars.
Identifying and understanding the origin and mechanism of carbon enhancement in these stars may therefore provide insight into the nature of the first stars, exotic supernovae, and star formation in the early universe.

A wide range of possible origins for carbon enhancement in CEMP--no stars have been proposed, each of which has different implications for the nature of Pop III stars and chemical enrichment.
Stars like \eri\ can help us to constrain and differentiate between these different scenarios.
In particular, so-called faint supernovae of Pop III stars have long been suggested as a possible progenitor of CEMP--no stars \citep{umeda_first-generation_2003,beers_discovery_2005,aoki_carbon-enhanced_2007,yoon_observational_2016,chiaki_seeding_2020}.
In a faint SN, the outer shells of lighter elements like C are ejected, while inner layers where heavier elements were synthesized fall back onto the remnant \citep{umeda_first-generation_2003}.
A variation is the `mixing-and-fallback' SN, which also invokes a strong degree of mixing in addition to a small ejecta mass \citep{umeda_first-generation_2003}.

Additionally, \cite{yoon_origin_2019} point out that there appear to be at least two sequences of CEMP--no stars in the A(C)\footnote{$A(X)=\log\epsilon(X)=\log (N_{x}/N_{H}) + 12$, where $N_{x}$ and $N_{H}$ are the number density of an element $X$ and Hydrogen}-[Fe/H] plane at moderate and low A(C), which they attribute to enrichment dominated by metal-free SNe and enrichment dominated by normal CCSNe respectively.
Spinstars (rapidly rotating massive stars) are also thought to produce excessive C, as well as N and O \citep{meynet_early_2006, jeena_rapidly_2023}.
Furthermore, ISM enrichment by AGB winds could be a dominant contributor even at very low metallicities \citep{sharma_chemical_2019,rossi_understanding_2023}.
We note that this mechanism is likely to produce stars that are at least $s-$process normal, which generally disfavors AGB wind contributions to carbon enhancement in UFDs.
Lastly, \cite{hartwig_formation_2019} suggest inhomogeneous metal mixing from normal CCSNe as an explanation for some fraction of CEMP--no stars. 
Carbon-rich gas cools more readily than carbon-normal gas, naturally favoring the formation of CEMP stars.

With these myriad formation mechanisms in mind, how should \eri's carbon enhancement be interpreted?
As discussed, the progenitor fits in Section \ref{sec:progenitors} are inconclusive.
A faint (low energy) SN is certainly within the broad range of possible progenitors, although it is not an obviously better fit to the observed abundance pattern than another class of progenitor.
The low dilution mass of the faint SN models also suggests that one such SN does not produce enough metals to plausibly be the sole contributor. 
Even with additional abundances or smaller uncertainties, uncertainties and degeneracies in the models themselves make it challenging to confidently discriminate between SN progenitor models, except in extreme cases \citep{placco_metal-poor_2015}. 

As it is challenging to differentiate between individual supernova models, it can be more helpful to use them in aggregate and consider specific abundance spaces that are more likely to be due to mono- versus multi-enrichment, regardless of the specific energy or mass of the progenitor \citep[e.g.,][]{hartwig_descendants_2018, vanni_chemical_2024}.
For example, the [Mg/C] ratio has been suggested as diagnostic of mono- versus multi-enrichment (\citealt{hartwig_descendants_2018}; see also \citealt{purandardas_chemical_2021, hansen_chemical_2024}).
In the right panel of Figure \ref{fig:c_fe}, we show the approximate distribution of multi- and mono-enriched stars in \citet{hartwig_descendants_2018}.
\cen\ lies solidly in the region expected to contain multi-enriched stars, as expected.
\eri\ and the other CEMP stars in the sample tend to have lower [Mg/C], which places them in the tails of the multi-enriched distribution.

\begin{figure*}
\centering
\includegraphics[width=\textwidth]{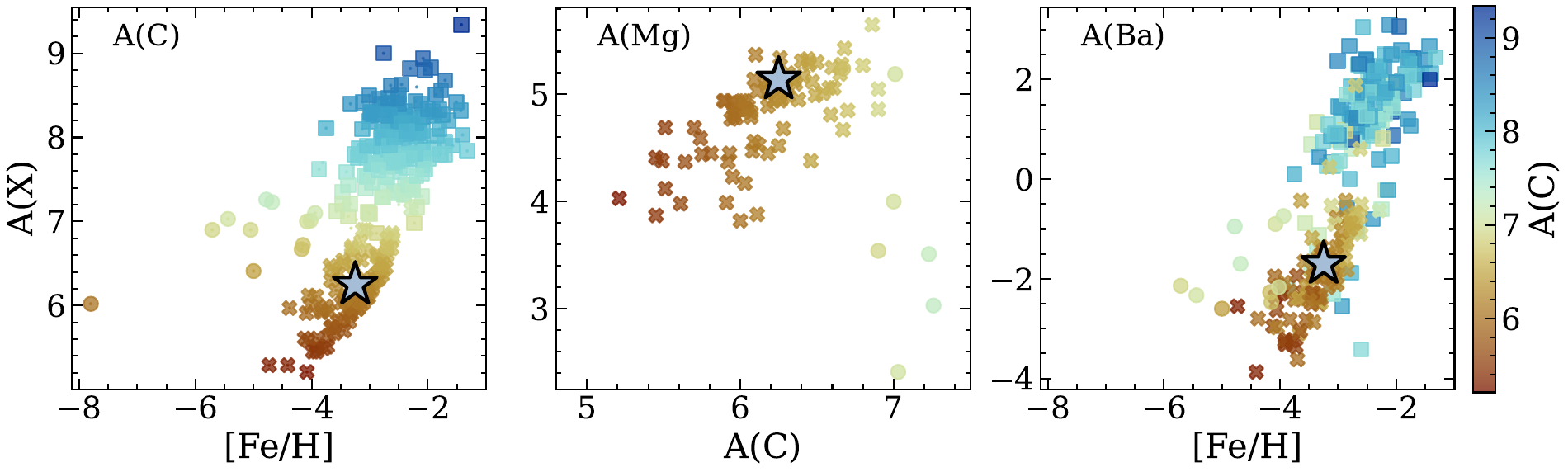}
\caption{A(X)-[Fe/H] for a sample of carbon-enhanced metal-poor stars \citep{yoon_observational_2016}; their Groups I, II, and III are denoted by squares, Xs, and circles respectively. \eri\ is marked as a light blue star. (Left) A(C)-[Fe/H] plane. \eri\ lies with Group II, the low-carbon group. The low-carbon and metallicity-independent moderate-carbon band (Group III, at about A(C) = 7) are CEMP--no stars, while the high-carbon group are typically CEMP--$s/i$ stars. (Center) A(Mg)-A(C) plane. The low-carbon group and moderate-carbon band separate out somewhat more readily in this space. The location of \eri\ in this plane suggests its carbon-enhancement has a common origin with the low-carbon group. (Right) A(Ba)-[Fe/H]. As a CEMP--no star, \eri\ lies at low A(Ba).\label{fig:cemp}}
\end{figure*}

We can also consider the A(C)-[Fe/H] plane. 
As Figure \ref{fig:cemp} shows, \eri\ lies in a similar region in the A(C)-[Fe/H] and A(Mg)-A(C) planes as the low carbon sequence from \cite{yoon_observational_2016} (their Group II), indicative of a common origin with the stars in this sequence.
The positive correlations in the A(C)-[Fe/H] and A(Mg)-A(C) planes for this subgroup of CEMP stars suggest progressive chemical evolution, rather than a stochastic source, as the origin of their carbon enhancement.
\cite{rossi_understanding_2023} finds that in the region of the A(C)-[Fe/H] plane that \eri\ occupies, dominant contributions from Pop III/II SNe are about equally likely.
This region is also enriched by AGB stars in their model, but high A(Mg) and low neutron-capture abundances both disfavor intrinsic AGB enrichment.
Lastly, \citet{rossi_understanding_2023} find that $-0.5\lesssim\textrm{[C/Mg]}\lesssim1.0$ suggests dominant Pop III enrichment (but not specifically mono-enrichment); \eri\ has $\mathrm{[C/Mg]} = 0.29\pm0.13$ (Figure \ref{fig:c_fe}).

We are unable to confidently rule out any particular enrichment scenario, but we favor multi-enrichment dominated Pop III SNe. 
First, the degree of carbon enhancement in \eri\ can be produced by mono- or multi-enrichment by a range of Pop III progenitors \citep[e.g.,][]{rossi_hidden_2024}, but it is nonetheless a relatively fragile signal, easily washed out by Pop II enrichment \citep{ji_preserving_2015, hartwig_descendants_2018}.
Second, \eri\ is metal-poor enough to allow for mono-enrichment, but not metal-poor enough to definitively favor it, and the fits to supernova progenitor models and other chemical diagnostics generally disfavor mono-enrichment.
Relatively low A(C) and [C/Fe]$< 2.0$ specifically point away from faint SNe specifically, which tend to produce stronger carbon enrichment \citep[e.g.,][]{yoon_origin_2019, jeon_role_2021, rossi_understanding_2023}.
We also cannot rule out contributions from spinstars.

\subsection{Chemical evolution of Eri~IV}\label{sec:mdf}
The metallicity distribution function of Eri~IV also provides some insight into the abundances of \eri.
Faint SNe under-produce iron relative to normal CCSN, so early enrichment dominated by faint SN in Eri~IV would produce a more pronounced metal-poor tail compared to enrichment dominated by normal CCSN \citep{komiya_are_2020,rossi_hidden_2024}.
The apparently right-skewed (left-leaning) metallicity distribution function of Eri~IV is inconsistent with this picture, especially when considering that \eri\ is the most metal-poor star identified in the system.
Of course, this does not preclude contributions from faint SNe, but the left-skewed metallicity distribution does suggest that faint SNe were not likely to be the only or even dominant source of metals in Eri~IV at early times.

Eri~IV also has an extremely low mean metallicity ($<$[Fe/H]$>= -2.87^{+0.08}_{-0.07}$).
This points to an especially abbreviated star formation history, which we might expect to retain the signatures of exotic supernovae, so it is perhaps surprising that its most metal-poor known star and its metallicity distribution do not.
In fact, the abbreviated, inefficient star formation characteristic of most UFDs has been proposed to explain the difference in fractions of CEMP stars between UFDs and more massive dSphs \citep{salvadori_carbon-enhanced_2015,yoon_origin_2019}. 
While it is unclear whether the CEMP fraction as a function of metallicity differs between UFDs and dSphs \citep{salvadori_carbon-enhanced_2015,lucchesi_extremely_2024}, extended star formation results in progressive iron enrichment that dilutes the signature of Pop III SNe (and also results in far fewer extremely metal-poor stars) \citep{salvadori_carbon-enhanced_2015,chiti_detection_2018,yoon_origin_2019}.

It is important to point out that uncertainty around the origin of \eri's enrichment is affected not only by uncertainties around the origin of carbon enhancement in general, but also by the particular chemical evolution of Eri~IV.
Mechanisms like outflows, gas infall, and mixing affect the dilution of elements in the ISM \citep{andrews_inflow_2017}; we also have no information about its star formation history.
To better understand the abundances of \eri\ and its context in the chemical evolution of Eri~IV, detailed abundances of more stars in the system are necessary.
Unfortunately, most of the other known member stars push the magnitude limits of current high-resolution spectrographs.
For these fainter stars, some abundances such as Mg and Ca can likely be measured from lower resolution spectra \citep[e.g.,][]{bonifacio_pristine_2025}, so this may be a viable route for future work.

\section{Conclusion}\label{sec:conclusion}

We present detailed chemical abundances for one star each in the ultra-faint dwarf galaxies Eri~IV and Cen~I using high-resolution Magellan/MIKE spectra. 
We find that the abundances of these stars are generally consistent with other UFDs.
In particular, both \eri\ and \cen\ have low neutron-capture abundances, which is characteristic of UFDs and could be a consequence of stochastic enrichment and/or star formation on short timescales compared to neutron-capture enrichment.
As \eri\ is an extremely metal-poor star, we also fit yields from metal-free supernovae models to its abundances.
We find a slight preference for a low-mass, normal-energy progenitor, but overall the fit to these models is poor.

Lastly, we consider the carbon enrichment of \eri, which is a CEMP--no star.
When considering the census of CEMP stars and the abundances of \eri, we argue that enrichment by a single faint SN is unlikely, but enrichment dominated by Pop III supernova is possible.
The left-skewed metallicity distribution function of Eri~IV is also inconsistent with enrichment strongly dominated by faint SNe.
However, this conclusion is tentative, and complicated by not only uncertainties in the origins of CEMP stars, but also uncertainties in the chemical evolution of Eri~IV.

The difficulty in interpreting the carbon enhancement of \eri\ in isolation highlights the need for chemical abundances of more stars in this system.
A couple of the brighter stars in Eri~IV (as well as several stars in Cen~I) are viable targets for high-resolution spectroscopy, while lower-resolution spectroscopy could provide abundance measurements for a few elements for the fainter stars.
Chemical abundances of stars across the metallicity distribution would provide valuable constraints on the enrichment history of this system and provide context for \eri.

\begin{acknowledgments}
The DELVE project is partially supported by Fermilab LDRD project L2019-011, the NASA Fermi Guest Investigator Program Cycle 9 grant 91201, and the U.S. National Science Foundation (NSF) under grants AST-2108168 and AST2307126.

This paper includes data gathered with the 6.5 meter Magellan Telescopes located at Las Campanas Observatory, Chile.

This work has made use of data from the European Space Agency (ESA) mission Gaia (https://www. cosmos.esa.int/gaia), processed by the Gaia Data Processing and Analysis Consortium (DPAC, https://www. cosmos.esa.int/web/gaia/dpac/consortium). Funding for the DPAC has been provided by national institutions, in particular the institutions participating in the Gaia Multilateral Agreement. 

M.E.H and T.S.L acknowledge the financial support of the Data Sciences Institute at the University of Toronto. G.E.M. and T.S.L. acknowledge financial support from Natural Sciences and Engineering Research Council of Canada (NSERC) through grant RGPIN-2022-04794. G.E.M. acknowledges support from an Arts \& Science Postdoctoral Fellowship at the University of Toronto. The Dunlap Institute is funded through an endowment established by the David Dunlap family and the University of Toronto.

W.C. gratefully acknowledges support from a Gruber Science Fellowship at Yale University. This material is based upon work supported by the National Science Foundation Graduate Research Fellowship Program under Grant No. DGE2139841. Any opinions, findings, and conclusions or recommendations expressed in this material are those of the author(s) and do not necessarily reflect the views of the National Science Foundation.

\end{acknowledgments}


\bibliography{heiger_chemical_abundances_in_eriIV_cenI}{}
\bibliographystyle{aasjournalv7}



\end{document}